\documentclass[12pt]{article}
\usepackage{epsfig}
\usepackage{psfrag}
\usepackage{epsfig}

\begin{document}
\thispagestyle{empty}
\noindent\hspace*{\fill} FAU-TP3-04/02 \\
\noindent\hspace*{\fill} hep-th/0402014 \\
\noindent\hspace*{\fill} \today \\

\begin{center}\begin{Large}\begin{bf} 
Phase Diagram of the Gross-Neveu Model: Exact Results and Condensed Matter Precursors
\\ 
\end{bf}\end{Large}\vspace{.75cm}
\vspace{0.5cm} Oliver Schnetz, Michael Thies and Konrad Urlichs\footnote{Electronic addresses:
thies@theorie3.physik.uni-erlangen.de,  konrad@theorie3.physik.uni-erlangen.de}\\ Institut f\"ur Theoretische Physik III \\
Universit\"at Erlangen-N\"urnberg, Erlangen, Germany\\
\vspace{1cm}\baselineskip=35pt
\end{center}
\date{\today}
\begin{abstract}
\noindent
Recently the revised phase diagram of the (large $N$) Gross-Neveu model in 1+1 dimensions
with discrete chiral symmetry has been determined numerically. It features three phases,
a massless and a massive Fermi gas and a kink-antikink crystal.
Here we investigate the phase diagram by analytical means, mapping the Dirac-Hartree-Fock
equation onto the non-relativistic Schr\"odinger equation with the (single gap) Lam\'e potential.
It is pointed out that mathematically identical phase diagrams appeared in the condensed matter
literature some time ago in the context of the Peierls-Fr\"ohlich model and ferromagnetic
superconductors.
\end{abstract}

\newpage

\vskip 0.5cm
\section{Introduction}

The Gross-Neveu (GN) model in 1+1 dimensions \cite{1} is probably the simplest interacting fermionic
field theory one can write down,
\begin{equation}
{\cal L} = \bar{\psi}^{(i)} {\rm i} \gamma^{\mu} \partial_{\mu} \psi^{(i)}
+ \frac{1}{2} g^2 \left( \bar{\psi}^{(i)} \psi^{(i)} \right)^2
\label{i1}
\end{equation}
(here, $i=1...N$ is a flavor index, and the model is defined through the 't Hooft limit $N\to \infty$, $Ng^2=$ const.\ \cite{2}).
Yet, as evidenced by more than 900 citations in the hep-archive to date, this toy model has turned out
to be quite valuable for a variety of physics questions (for a recent review and a discussion of
subtleties associated with low dimensions, see Ref.\ \cite{3}).
If one solves it via semi-classical methods following the original 1974 paper, a surprising number of phenomena
of interest to strong interaction physics unfold. Already in Ref.\ \cite{1} asymptotic freedom, spontaneous breaking of the
discrete chiral symmetry $\psi \to \gamma_5 \psi$,
dynamical fermion mass generation and a scalar $q\bar{q}$ bound state (the $\sigma$-meson)
were demonstrated. The version with continuous chiral symmetry, the two-dimensional
Nambu--Jona-Lasinio model \cite{4}
(NJL$_2$) possesses an additional massless $\pi$-meson. Shortly afterwards, massive kink \cite{6} and
kink-antikink \cite{7} type baryons were derived analytically in the discrete chiral model,
whereas the NJL$_2$ model features massless baryons \cite{8,8b}.
Promptly it was shown that chiral symmetry gets restored at finite temperature in a 
second order phase transition \cite{8a,8c}.

In contrast to these early works, it took a rather long and winding
road to determine the full phase structure of the GN model at finite chemical potential and temperature. 
In 1985, the first phase diagram was proposed \cite{9}, followed by a number of works elaborating on it (see e.g.\ \cite{10,10a}).
Its prominent features were two phases (massive and massless quarks), separated by a line of first
and second order transitions meeting at a tricritical point. In a density-temperature plot, a mixed phase would appear in the
region of low density and temperature. It can be pictured as droplets of chirally restored vacuum
containing extra quarks, embedded in the symmetry broken vacuum --- reminiscent of the MIT bag model \cite{11}.
At the first order phase transition, the droplets fill all space \cite{3}.

This scenario as well as the underlying phase diagram were
believed to hold for both variants of the GN model (with discrete and continuous chiral symmetry) \cite{10a}. 
Only in 2000 it was noticed that the phase diagram was hard to reconcile with the known baryon spectrum.
In the case of the NJL$_2$ model, a thermodynamically more stable solution of the mean field equations turned out to
be one where the chiral condensate assumes a helical form, namely a circle in
the ($\bar{\psi}\psi, \bar{\psi}{\rm i}\gamma_5 \psi$)
plane superimposed with a uniform translation along the $x$ axis (a ``chiral spiral" \cite{14}).
Since the winding number is equal to baryon number,
one can regard this structure as a caricature of a Skyrme crystal \cite{13}, a point of view emphasized in Ref.\ \cite{14}.
This helical order parameter was subsequently confirmed in Ref.\ \cite{15}, the (substantially) modified phase diagram of the
NJL$_2$ model is discussed in Ref.\ \cite{3}.

As the bosonization technique based on the massless boson is unavailable in the GN model with discrete
chiral symmetry, it took somewhat longer to construct a satisfactory phase diagram for this model.
Following a variational calculation which clearly showed the existence of a crystal ground
state at $T=0$ and any finite density \cite{15a},
the full phase diagram was obtained in Ref.\ \cite{16}, based on a numerical solution of the Dirac-Hartree-Fock equations.
The main new feature is the appearance of a third phase, a kink-antikink crystal, in addition to the two previously 
known homogeneous phases. It supersedes the mixed phase in the old phase diagram.
All transitions are second order, and
the tricritical point gets transmuted into a Lifshitz point characteristic for the transition from
an inhomogeneous to a homogeneous ordered phase in condensed matter physics \cite{17}.
The mechanism which drives the spontaneous breakdown of translational invariance at finite density was
identified as Overhauser effect \cite{18} with gap formation at the Fermi surface. More recently,
guided by these results, the analytic form of the self-consistent scalar potential at $T=0$ has been
guessed correctly and the self-consistency of the crystal ground state could be established analytically \cite{19}.
Many of the previous numerical results can now be written down in a concise, analytical way.
Instrumental for this solution was the fact that by using an appropriate scalar potential expressed in
terms of Jacobi elliptic functions, the Dirac-Hartree-Fock equation could be mapped onto
a Schr\"odinger equation with the 
Lam\'e potential which was solved analytically long time ago \cite{20}.

In the present paper, we extend the work of Ref.\ \cite{19} to finite temperature. It turns out that the ansatz for
the self-consistent potential used in \cite{19} is general enough to encompass the self-consistent potential
at any temperature and chemical potential. Apart from the obvious benefit of analytic insight into
the phase structure of the GN model, our results have proven helpful in uncovering some 
striking relationship between the
GN model and certain problems in condensed matter physics, notably in the context of superconductivity.

This paper is organized as follows: 
In Sect.\ \ref{section2}, we present our two-parameter ansatz for the scalar potential and the solution of the
Hartree-Fock-Dirac
equation. In Sect.\ \ref{section3}, we minimize the grand canonical potential and show self-consistency at finite
temperature and
chemical potential. We also discuss how to compute thermodynamic observables. In Sect.\ \ref{section4}, we map out the
phase boundaries by appropriate series expansions, study the $T=0$ limit and derive an effective action of Ginzburg-Landau
type near the multicritical point. 
Finally in Sect.\ \ref{section5} we point out the close
relationship between the phase diagram of the GN model and quasi-one-dimensional condensed matter problems.


\section{Ansatz for the scalar potential and solution of the Dirac-Hartree-Fock equation}
\label{section2}
\begin{figure}[h]
\begin{center}
\begin{psfrags}
\psfrag{omega}{$\omega$}
\psfrag{P}{$p$}
\psfrag{kappa}{$\displaystyle \sqrt{1-\kappa^2}$}
\psfrag{Left}{$\displaystyle -\frac{\pi}{2\mathbf{K}}$}
\psfrag{Right}{$\displaystyle \frac{\pi}{2\mathbf{K}}$}
\psfrag{lower band}{lower band}
\psfrag{upper band}{upper band}
\epsfig{file=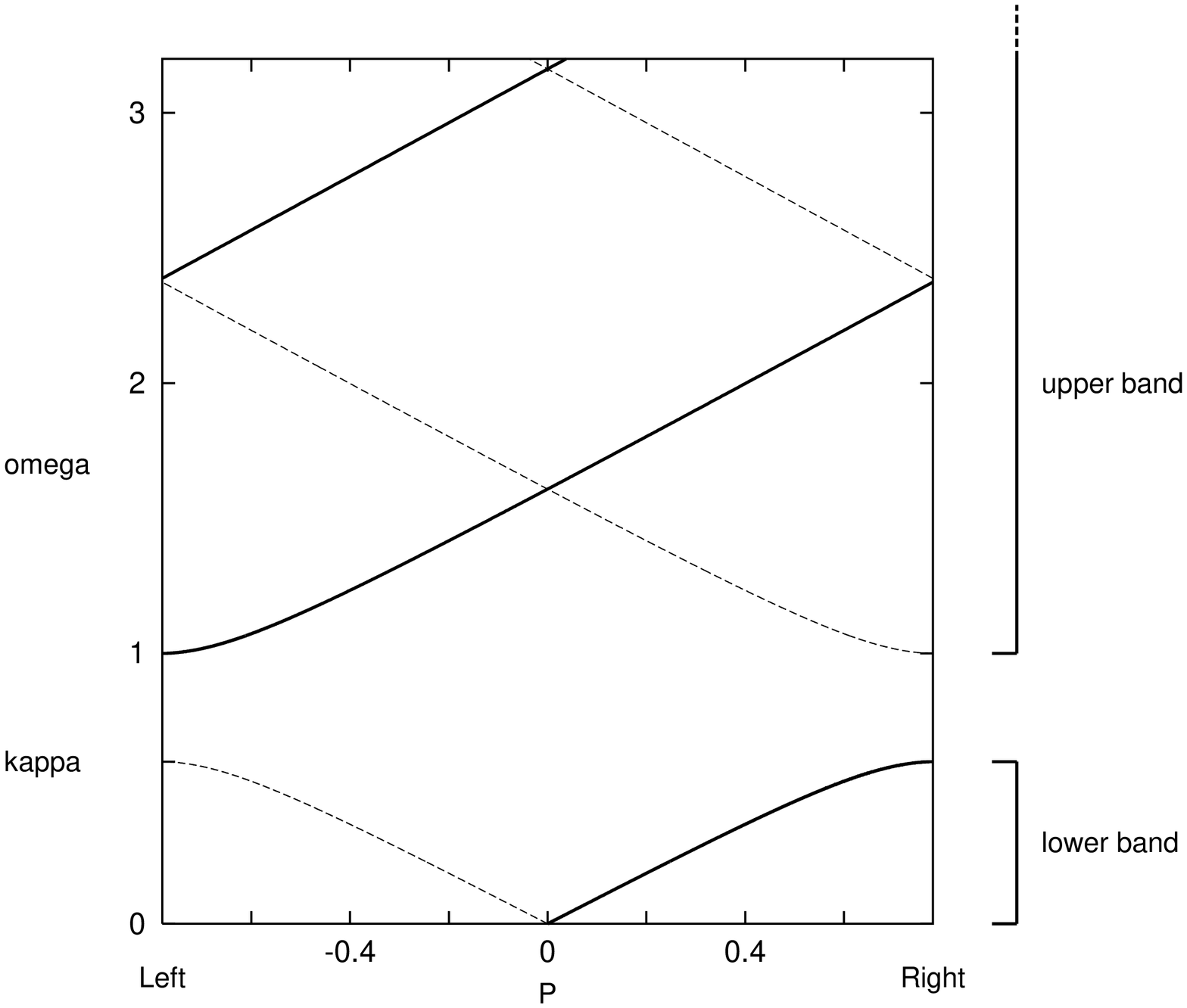,width=10cm, angle=-90}
\caption{The dispersion relation~(\ref{A19a}) for $\kappa=0.8$. The Bloch momentum $p$ is projected onto the
Brillouin zone ranging from $-\pi/2\mathbf{K}$ to $+\pi/2\mathbf{K}$. The bold line marks the branch for which the slope 
of the
dispersion relation is positive. Notice the symmetry $p \to -p$.}
\label{FIG1}
\end{psfrags}
\end{center}
\end{figure}

We apply the relativistic Hartree-Fock approximation which becomes exact in the large-$N$ limit (see \cite{3} for details). 
The starting
point for our discussion is the Dirac-Hartree-Fock equation,
\begin{equation}
\left( \gamma^5 \frac{1}{\rm i} \frac{\partial}{\partial x} + \gamma^0 S(x)\right)
\psi(x) = E \psi(x) \, ,
\label{A2}
\end{equation}
with the following choice of $\gamma$-matrices,
\begin{equation}
\gamma^0 =-\sigma_1 \, , \quad \gamma^1 ={\rm i}\sigma_3 \, , \quad \gamma^5 = \gamma^0
\gamma^1 =-\sigma_2 \, .
\label{A3}
\end{equation}
In the Gross-Neveu model with discrete chiral symmetry, $S(x)$ is real.
In terms of the upper and lower spinor components $\phi_{\pm}$
the Dirac equation consists of two coupled equations
\begin{equation}
\mp \left( \frac{\partial}{\partial x} \pm S \right)\phi_{\pm} = E \phi_{\mp}
\label{A4}
\end{equation}
which can be decoupled by squaring,
\begin{equation}
\left( - \frac{\partial^2}{\partial x^2} \mp \frac{\partial S}{\partial x} + S^2 \right)
\phi_{\pm} = E^2 \phi_{\pm} \, .
\label{A5}
\end{equation}
Eqs.\ (\ref{A2}-\ref{A5}) fall into the pattern of supersymmetric (SUSY) quantum mechanics. As ansatz for $S$ we
choose the
superpotential of the Lam\'{e} potential \cite{21},
\begin{equation}
S(x)=A \kappa^2 \frac{{\rm sn}(Ax, \kappa) {\rm cn}(Ax,\kappa)}
{{\rm dn}(Ax, \kappa)} \, ,
\label{A6}
\end{equation}
where $\mathrm{sn}$, $\mathrm{cn}=\sqrt{1-\mathrm{sn}^2}$, and $\mathrm{dn}=\sqrt{1-\kappa^2 \mathrm{sn}^2}$
are Jacobi elliptic functions to the modulus~$\kappa$ \cite{22}. We introduce dimensionless variables
\begin{equation}
\xi =Ax,\quad \tilde{S}(\xi)=S(x)/A, \quad \omega=E/A
\label{dimensionless}
\end{equation}
and find that $\tilde{S}$ has period $2\mathbf{K}$, where $\mathbf{K}=\mathbf{K}[\kappa]$ is the complete elliptic integral
of the first
kind. Moreover the scalar potential satisfies the symmetry condition
\begin{equation}
\tilde{S}(\xi + \mathbf{K}) = - \tilde{S}(\xi) \, .
\label{A6a}
\end{equation}
With the ansatz~(\ref{A6}), the second order equation~(\ref{A5}) becomes
\begin{equation}
\left( - \frac{\partial^2}{\partial \xi^2} + 2 \kappa^2 {\rm sn}^2(\xi , \kappa) \right)
\phi_+ (\xi) = (\omega^2+\kappa^2) \phi_+ (\xi)
\label{A11}
\end{equation}
the simplest case of the Lam\'{e} equation (with a single gap). The solutions are well known \cite{20,23},
\begin{equation}
\phi_+(\xi)=\mathcal{N} \frac{\mathrm{H}(\xi+\alpha)}{\Theta(\xi)}\mathrm{e}^{- \mathrm{Z}(\alpha) \xi}
\label{lamesol}
\end{equation}
where $\mathrm{H}$, $\Theta$ and $\mathrm{Z}$ are the Jacobi eta, theta and zeta function, 
respectively \cite{22}. (Regretfully there exist different conventions for the argument of the Jacobi zeta function. We choose
the convention with $\mathrm{Z}(\mathbf{K}) = 0$). The parameter $\alpha$ is related to the (reduced) energy $\omega$ and
Bloch momentum $p$ by
\begin{equation}
|\omega|=\mathrm{dn}(\alpha,\kappa), \quad p = -\mathrm{i}\mathrm{Z}(\alpha) + \frac{\pi}{2\mathbf{K}} \, .
\label{A19a}
\end{equation}
Note that $\alpha$ and $-\alpha$ supply the two independent solutions for a given energy eigenvalue $\omega$. The
condition for real valued energy and momentum leads to two energy bands $\omega^2=0\ldots (1-\kappa^2)$,
$p=0 \ldots \pi/2\mathbf{K}$ and $\omega^2=1\ldots\infty$, $p=\pi/2\mathbf{K}\ldots\infty$. The lower band is
parametrized by $\alpha = \mathbf{K}+\mathrm{i}(\mathbf{K}'-\eta)$, the upper band by $\alpha = \mathrm{i}\eta$, where in both
cases $\eta = 0\ldots \mathbf{K}'\equiv\mathbf{K}[\sqrt{1-\kappa^2}]$. From Eq.\ (\ref{A19a}) we calculate
($\mathbf{E}=\mathbf{E}[\kappa]$ is the complete elliptic integral of the second kind)
\begin{equation}
\frac{\mathrm{d}p}{\mathrm{d}{\omega}} = \pm \frac{\omega^2-\mathbf{E}/\mathbf{K}}{\sqrt{(\omega^2-1
+\kappa^2)(\omega^2-1)}} \, ,
\label{dispersion}
\end{equation}
where the plus sign refers to the upper band, the minus sign to the lower band. We chose the sign of $\alpha$ in such a way
 that the slope of the dispersion relation is positive in both bands. The eigenfunction corresponding to $-\alpha$ has
 quasi-momentum $\pi/\mathbf{K}-p$. The full dispersion relation is plotted in Fig.\ \ref{FIG1}. \par
With $\phi_+$ given, $\phi_-$ follows from the Dirac equation~(\ref{A4}),
\begin{equation}
\phi_-=-\frac{1}{\omega}\left(\frac{\partial}{\partial \xi}+\tilde{S}\right)\phi_+ \, .
\end{equation}
More explicitly, from Eqs.\ (\ref{A4}) and (\ref{A6a}) we can read off that a simultaneous translation $\xi \to \xi
+\mathbf{K}$ and a discrete chiral transformation $\psi \to \gamma^5\psi$ leaves the Dirac equation invariant. Thus
$\gamma^5\psi(\xi+\mathbf{K})\propto \psi(\xi)$, or equivalently $c\phi_+(\xi)=\phi_-(\xi+\mathbf{K})$,
$c\phi_-(\xi)=-\phi_+(\xi+\mathbf{K})$ for some constant $c$. By iteration we find
\begin{equation}
c^2\phi_+(\xi)=-\phi_+(\xi+2\mathbf{K}) =-\mathrm{e}^{\mathrm{i}p2\mathbf{K}}\phi_+(\xi) \, .
\end{equation}
Comparison with~(\ref{lamesol}) determines $c$ up to a sign ambiguity which can be resolved by an explicit calculation, 
\begin{equation}
\phi_-(\xi) = -\mathrm{sgn}(\omega) {\rm e}^{\mathbf{K} Z(\alpha)} \phi_+(\xi+\mathbf{K}) \, .
\label{b11a}
\end{equation}
We still have to determine the normalization factor ${\cal N}$ in Eq.\ (\ref{lamesol}). In a continuum normalization, the spatially 
averaged fermion density should be normalized to 1,
\begin{equation}
1 = \langle \psi^{\dagger}\psi \rangle = \frac{1}{2\mathbf{K}} \int_0^{2 \mathbf{K}}{\rm d}\xi
\, 2|\phi_+|^2 \, .
\label{A16}
\end{equation}
Inserting (\ref{lamesol}) and using an addition theorem for $\mathrm{H}$ (note that $\mathrm{H}(x+y)\mathrm{H}(x-y)/\Theta^2(x)
\Theta^2(y)$ is elliptic in $x$ and $y$ and thus determined up to an additive constant by its pole structure)
\begin{equation}
\frac{\mathrm{H}(x+y)\mathrm{H}(x-y)}{\Theta^2(x)\Theta^2(y)}=\frac{\pi}{2\mathbf{K}\kappa\sqrt{1-\kappa^2}}\left(\mathrm{dn}^2
(y,\kappa)-\mathrm{dn}^2(x,\kappa)\right)
\label{A21}
\end{equation}
leads to
\begin{equation}
|\phi_+|^2 = |{\cal N}|^2 \frac{\pi\Theta^2(\alpha)}{2\mathbf{K}\kappa\sqrt{1-\kappa^2} }\left|\mathrm{dn}^2(\alpha,\kappa)-
\mathrm{dn}^2(\xi,\kappa)\right| \, .
\label{A22}
\end{equation}
The spatial average is easily computed. We find 
\begin{equation}
|\mathcal{N}|^2=\frac{\mathbf{K}\kappa\sqrt{1-\kappa^2}}{\pi \Theta^2(\alpha)|\mathrm{dn}^2(\alpha,\kappa)-\mathbf{E}
/\mathbf{K}|} \, .
\
\end{equation}
With Eq.\ (\ref{A19a}) we can simplify the expression for $|\phi_+|^2$ to
\begin{equation}
|\phi_+|^2 = \frac{1}{2} \frac{\omega^2-\mathrm{dn}^2(\xi,\kappa)}{\omega^2-\mathbf{E}/\mathbf{K}} \, .
\label{20a}
\end{equation}
Let us now calculate $\bar{\psi}\psi$ and $\psi^\dagger\psi$, the scalar and baryon densities
for a single orbit. We find
\begin{equation}
\bar{\psi}\psi=-(\phi_+^*\phi_-+\phi_-^*\phi_+) = \frac{1}{\omega}
(\partial_\xi+2\tilde{S})|\phi_+|^2
\end{equation}
and with Eq.\ (\ref{20a}) 
\begin{equation}
\bar{\psi} \psi =\frac{\omega}{\omega^2 -\mathbf{E}/\mathbf{K}}\tilde{S} \, .
\label{scalardensity}
\end{equation}
Notice that every orbit has the $\xi$-dependence of the Lam\'{e} superpotential. On the other hand Eq.\ (\ref{b11a}) implies
$\psi^\dagger\psi =|\phi_+(\xi)|^2+|\phi_+(\xi+\mathbf{K})|^2$. Using Eq.\ (\ref{20a}) and the 
addition theorem for $\mathrm{dn}(\xi+\mathbf{K},\kappa)$ leads to
\begin{equation}
\psi^\dagger\psi=\frac{\omega^2-1+\kappa^2/2+\tilde{S}^2/2}{\omega^2-\mathbf{E}/\mathbf{K}} \, .
\label{baryondensity}
\end{equation}
>From the normalization condition $\langle \psi^\dagger \psi\rangle =1$ we deduce the spatial average
\begin{equation}
s^2\equiv \langle \tilde{S}^2\rangle =2-\kappa^2-2\mathbf{E}/\mathbf{K} \, .
\label{Sspatialav}
\label{b11c}
\end{equation}
This completes the preparations needed for calculating the grand canonical potential density.


\section{The grand canonical potential}
\label{section3}
In order to determine the phase diagram of the GN model, one has to minimize the grand canonical potential in the space
of scalar potentials $S(x)$.
We will show that the two-parameter ansatz (\ref{A6}) provides a self-consistent solution so that it is
sufficient to minimize with respect to $A$ and $\kappa$. In the relativistic Hartree-Fock
approximation, the grand canonical potential density per flavor is given by (see \cite{3} and references therein)
\begin{equation}
\Psi= - \frac{1}{\beta \pi} \int_0^{\Lambda /2}{\rm d}q\, \ln
\left[ \left(1+{\rm e}^{-\beta (E-\mu)}\right) \left(1+{\rm e}^{\beta (E+\mu)}\right) \right]
+\frac{1}{2Ng^2\lambda} \int_0^{\lambda}{\rm d}x S^2(x)\, ,
\label{c1}
\end{equation}
Here, $\Lambda/2$ is the ultra-violet cutoff, $\lambda=2\mathbf{K}/A$ the spatial
period of $S(x)$, and $q=Ap$. We change to $\omega$ as integration variable, spelling out the contributions from the
 two energy bands. With Eqs.\ (\ref{dispersion}) and
(\ref{Sspatialav}) we find
\begin{equation}
\Psi = -\frac{A}{\beta \pi} \left(\hspace{-0.1cm}\int\limits_0^{\sqrt{1-\kappa^2}}\hspace{-0.3cm}\mathrm{d}\omega\, +
\int\limits_1^{\Lambda_{\omega}}\hspace{-0.2cm} \mathrm{d}\omega\right) \frac{\mathrm{d}p}{\mathrm{d}
\omega} \ln\left[\left(1+\mathrm{e}^{- \beta (A\omega- \mu)}\right)\left(1+\mathrm{e}^{\beta (A\omega+
\mu)}\right)\right] + \frac{A^2 s^2}{2Ng^2} \, .
\label{PSI}
\end{equation}
>From Eq.\ (\ref{A19a}) we extract the behavior of $\omega$ for large $p$ and find that $\Lambda_{\omega}=\omega
(q=\Lambda/2)=\Lambda/2A +As^2/\Lambda+\mathcal{O}(\Lambda^{-3})$. Due to the quadratic divergence, we have to
keep the $1/\Lambda$ term.\par
In the following it is convenient to combine the integral over both energy bands as well as over positive and negative energy
modes into the real part of a single integral in the complex plane. The path of integration is shifted slightly away from the
real axis. We find that $\mathrm{d}p/\mathrm{d}\omega$ defines a double cover of the complex
plane and choose the sheet by requiring $\mathrm{d}p/\mathrm{d}\omega>0$ for $\omega>1$. The gaps and
the sign-change in Eq.\ (\ref{dispersion}) match the behavior of $\mathrm{d}p/\mathrm{d}\omega$ in the complex plane.
Setting
\begin{equation}
a = \beta A, \quad \nu = \beta \mu \, ,
\end{equation}
we obtain
\begin{equation}
\pi\beta^2\Psi= \frac{\pi a^2 s^2}{2Ng^2}
-a\lim\limits_{\varepsilon\to 0} \, \mathrm{Re} \int\limits_{-\Lambda_{\omega} +\mathrm{i}\varepsilon}^{\infty+
\mathrm{i}\varepsilon} \mathrm{d}\omega \, \frac{\omega^2-\mathbf{E}/\mathbf{K}}{\sqrt{(\omega^2-1+\kappa^2)
(\omega^2-1)}}\ln\left(1+\mathrm{e}^{-a\omega +\nu}\right) \, .
\label{24b}
\end{equation}
The physical value of $\Psi$ for given $\beta$ and $\mu$ is determined by the minimum of $\Psi$ in the two-dimensional space
$(\kappa,A) \in (0\ldots 1, 0\ldots \infty)$. \par
It turns out to be convenient to minimize $\Psi$ in the directions given by two specific linear combinations of $\partial/\partial A$
and $\partial /\partial \kappa$, corresponding to $a \partial/\partial a + \mathbf{K} \partial/\partial \mathbf{K}$ and $ a\partial
/\partial a - s
\partial /\partial s$. Consequently we introduce
\begin{eqnarray}
F_1&=&\frac{1}{a^2(1-\mathbf{E}/\mathbf{K})}\left(a \frac{\partial}{\partial a}-\frac{1-\kappa^2}{1-\kappa^2-\mathbf{E}/\mathbf{K}}
\kappa \frac{\partial}{\partial \kappa}\right) \pi \beta^2\Psi \quad \mbox{ and} \nonumber \\
\label{B30}
F_2&=&\frac{1-\kappa^2-\mathbf{E}/\mathbf{K}}{a^2(1-\mathbf{E}/\mathbf{K})} \left( a\frac{\partial}{\partial a} - \frac{(2-\kappa^2-
2\mathbf{E}/\mathbf{K})(1-\kappa^2)}{(1-\kappa^2-\mathbf{E}/\mathbf{K})^2}\kappa \frac{\partial}{\partial \kappa}\right)\pi \beta^2
\Psi
\end{eqnarray}
where we have converted $\partial/\partial \mathbf{K}$ and $\partial/\partial s$ into the $\partial/\partial \kappa$ terms.
The calculation of $F_1$ and $F_2$ is straightforward (replace $\omega$ by $\omega /a$ in $\Psi$ before differentiation),
\begin{eqnarray}
\label{F1}
\hspace{-0.5cm} F_1 &=& {\frac{\pi}{Ng^2} -1 } + \displaystyle \frac{1}{a} \lim\limits_{\varepsilon\to 0}\mathrm{Re} \hspace{-0.4cm}
\int\limits_{-\frac{\Lambda\beta}{2a}+\mathrm{i}\varepsilon}^{\infty+\mathrm{i}\varepsilon} \hspace{-0.4cm} \mathrm{d}\omega\,
\left(\frac{\partial}{\partial \omega}\frac{\omega}{\sqrt{(\omega^2-1+\kappa^2)(\omega^2-1)}}\right) \ln
\left(1+\mathrm{e}^{-a\omega+\nu}\right) , \nonumber \\
\label{F2}
\hspace{-0.5cm} F_2 &=& \frac{1}{a} \lim\limits_{\varepsilon\to 0}\mathrm{Re} \hspace{-0.4cm} \int\limits_{-\infty+\mathrm{i}
\varepsilon}^{\infty+\mathrm{i}\varepsilon} \hspace{-0.4cm} \mathrm{d}\omega\, \left(\frac{\partial}{\partial \omega}
\frac{(1-\kappa^2)-\omega^2\mathbf{E}/\mathbf{K}}{\sqrt{(\omega^2-1+\kappa^2)(\omega^2-1)}}\right) \frac{1}
{\omega} \ln \left(1+\mathrm{e}^{-a\omega+\nu}\right) .
\end{eqnarray}
Note that the integral in $F_1$ is logarithmically divergent as $\Lambda\to\infty$, whereas the integral in $F_2$ is convergent.
Hence $F_2$ is a function of $a,\nu, \kappa$ only. In the range $0<\kappa<1$, $0<A$, minimizing $\Psi$ is equivalent to
$F_1=F_2=0$.
This defines the crystal phase. It exists only in a certain region in the $(T,\mu)$-plane (see \cite{16} and Sect.\ \ref{section4}).
In this region the value of $\Psi$
has to be compared with the (mathematically) co-existing value of $\Psi$ for the massless Fermi gas (corresponding to
$\kappa=0$ or $A=0$)
and the value of $\Psi$ for the massive Fermi gas (corresponding to $\kappa=1$). It has been checked numerically that the
crystal solution
(whenever it exists) is thermodynamically most stable. This result is confirmed by an analytical investigation
near the phase boundaries as well as at $T=0$,
cf. Sect.\ \ref{section4}.\par
One advantage of our particular choice of direction in which we minimized the potential [cf. Eq. (\ref{B30})] is
the fact that $F_1=0$ is equivalent to the self-consistency of our
ansatz. To show this we manipulate the thermal expectation value (note the sign change for negative $\omega$ in
Eq.\ (\ref{scalardensity}))
\begin{equation}
\label{A40}
\langle\bar{\psi}\psi\rangle_{\rm th} =
\frac{1}{\pi}\int\limits_0^{\Lambda/2}\mathrm{d}q \, \bar{\psi}\psi\left(\frac{1}{\mathrm{e}^{\beta(E-\mu)}
+1}-\frac{1}{\mathrm{e}^{\beta(-E-\mu)}+1}\right)
\end{equation}
in the way we reformulated $\Psi$. With Eqs.\ (\ref{dispersion}), (\ref{scalardensity}) we obtain
\begin{equation}
\label{A41}
\langle\bar{\psi}\psi\rangle_{\rm th} = \frac{1}{\pi}S(x)\lim\limits_{\varepsilon\to 0}\, \mathrm{Re}\hspace{-0.4cm}
\int\limits_{-\frac{\Lambda\beta}
{2a}+\mathrm{i}\varepsilon}^{\infty+\mathrm{i}\varepsilon} \hspace{-0.4cm} \mathrm{d}\omega \frac{\omega}
{\sqrt{(\omega^2-1+\kappa^2)(\omega^2-1)}}\frac{1}{(\mathrm{e}^{a\omega-\nu}+1)} \, .
\end{equation}
Applying a partial integration in Eq.\ (\ref{F1}) we can use $F_1$ to express $\langle\bar{\psi}\psi\rangle_{\rm th}$
as follows,
\begin{equation}
\langle\bar{\psi}\psi\rangle_{\rm th} = \frac{1}{\pi}S(x) \left(F_1 - \frac{\pi}{Ng^2}\right) .
\end{equation}
This in turn reduces to the self-consistency condition $S(x)=-Ng^2\langle\bar{\psi}\psi\rangle_{\rm th}$ for $F_1=0$.

In order to isolate the divergencies and to make expansions for $k\to 0$, $k\to 1$, and $a\to 0$ easier accessible, 
we cast $\Psi, F_1, F_2$
into a different form. The transformation amounts to changing from the integrals over the two energy bands 
to an integration over the
single gap. Here we simply give the results which can be verified by comparing the Taylor expansions of 
corresponding functions at $a=0$. Whereas the
results for the transformed expressions are trivial (cf. Eqs.\ (\ref{68}), (\ref{71})), some work is needed to 
expand Eqs.\ (\ref{24b}), (\ref{F2}) at $a=0$. The calculations are lengthy but straightforward if one proceeds as follows:
\begin{enumerate}
\item replace $\omega$ by $\omega /a$ as integration variable.
\item show that the expansion in $a=0$ can be derived from expanding the integrand.
\item use integration by parts twice to obtain the factor $[\ln (1+\mathrm{e}^{-\omega+\nu})]''=
\left(2\cosh \frac{\omega-\nu}{2}\right)^{-2}$.
This renders the integrals finite for $\Lambda \to \infty$.
\item The $a^0$- term can be converted into a sum and thus evaluated.
\item Using the residue theorem, the remaining terms can be converted into a sum over Matsubara frequencies
(the $a^2$ term needs
some extra attention).
\item The sum over the Matsubara frequencies can be evaluated yielding rational multiples of $\zeta (2k+1)$
(where $\zeta$ is the Riemann
zeta function and $k=1,2,\ldots$).
\item Comparison with the power series of $\mathrm{Im}\ln\Gamma\left(\frac{1}{2}+\frac{\mathrm{i}\nu}{2\pi}\right)$
 at $a=0$ allows
one to give the result in terms of $\mathrm{Im}\ln\Gamma$ and its derivatives.
\end{enumerate}
Proving the results is now equivalent to deriving Eq.\ (\ref{68}). \par
We reintroduce $A$ and $\mu$ to show the dependence on the basic parameters of our ansatz:
\begin{eqnarray}
\Psi& =& -\frac{\Lambda^2}{8\pi}-\frac{\mu\Lambda}{2\pi}-\frac{\pi}{6\beta^2}-\frac{\mu^2}{2\pi}+\frac{2A^2}{\pi^2} \int
\limits_0^{\pi/2}
\mathrm{d}\varphi \, \left(\Delta^2_{\varphi}-\frac{\mathbf{E}}{\mathbf{K}}\right) \nonumber \\
&& \left[\mathrm{Re}\frac{\pi}{\mathrm{i}\beta A \Delta_{\varphi}}\ln\frac{\Gamma\left(\frac{1}{2}+\frac{\mathrm{i}\beta}{2\pi}
(\mu
+A\Delta_{\varphi})\right)}{\Gamma\left(\frac{1}{2}+\frac{\mathrm{i}\beta}{2\pi}(\mu-A\Delta_{\varphi})\right)}-\ln
\frac{\Lambda\beta}{4\pi}
+\frac{\pi}{Ng^2}\right]
\end{eqnarray}
where
\begin{equation}
\Delta_{\varphi}=\sqrt{1-\kappa^2\sin^2(\varphi)} \, .
\end{equation}
The quadratically and linearly divergent terms (energy density $-\Lambda^2/8\pi$ and baryon density
$-\mu\Lambda/2\pi$ of the Dirac sea) are irrelevant and can simply be dropped. From now on, we will work in units where the
 dynamical fermion mass in the vacuum is 1. In these units the vacuum gap equation reads
\begin{equation}
\frac{\pi}{Ng^2} = \ln \Lambda \, .
\end{equation}
Invoking this equation, all remaining infinities are removed and we obtain the renormalized expression
\begin{equation}
\label{24i}
\pi \beta^2 \Psi_{\mathrm{ren}}= -\frac{\pi^2}{6} - \frac{\nu^2}{2} - \frac{a^2s^2}{2}\ln\frac{\beta}{4\pi} + 2 a \int\limits_0^{\pi/2}
\mathrm{d}\varphi \,
\left(\Delta_{\varphi} -\frac{\mathbf{E}}{\mathbf{K}\Delta_{\varphi}}\right)\mathrm{Im}\, \ln \frac{\Gamma\left(\frac{1}{2}+
\frac{\mathrm{i}}{2\pi}
(\nu+a\Delta_{\varphi})\right)}{\Gamma\left(\frac{1}{2}+\frac{\mathrm{i}}{2\pi}(\nu-a\Delta_{\varphi})\right)} \, .
\end{equation}
Similarly we obtain ($\psi=(\ln\Gamma)'$)
\begin{eqnarray}
\label{F1eq0}
F_{1,\mathrm{ren}} &=& - \ln\frac{\beta}{4\pi }+ \frac{1}{\pi} \int\limits_0^{\pi/2} \mathrm{d}\varphi \, \mathrm{Re} \left[
\psi\left(\frac{1}{2}+
\frac{\mathrm{i}}{2\pi}(\nu+a\Delta_{\varphi})\right)+ {(a \rightarrow -a)} \right] \, , \\
\label{F2eq0}
F_{2} &=& 2a\frac{\partial}{\partial a}\frac{1}{a}\int\limits_0^{\pi/2}\mathrm{d}\varphi \, \left(\frac{1-\kappa^2}{\Delta^3_{\varphi}}-
\frac{\mathbf{E}}{\mathbf{K} \Delta_{\varphi}}\right)\mathrm{Im}\ln \frac{\Gamma\left(\frac{1}{2}+\frac{\mathrm{i}}{2\pi}(\nu+a
\Delta_{\varphi})\right)}{\Gamma\left(\frac{1}{2}+\frac{\mathrm{i}}{2\pi}(\nu-a\Delta_{\varphi})\right)} \, .
\end{eqnarray}
We immediately see that $\Psi_{\mathrm{ren}}(\beta,a^2,\nu^2,\kappa^2)$, $F_{1,\mathrm{ren}}(\beta,a^2,\nu^2,\kappa^2)$
and $F_{2}(a^2,\nu^2,\kappa^2)$ are even, analytical functions in $a$, $\nu$ and $\kappa$.
A practical procedure to solve $F_{1,\mathrm{ren}}=F_2=0$ is as follows:
\begin{enumerate}
\item Choose values for two out of the three parameters $a=0\ldots\infty$, $\nu=0\ldots\infty$, $\kappa=0\ldots 1$.
\item Solve $F_2=0$ (Eq.\ (\ref{F2eq0})) for the third parameter. This requires finding the zeros of a function of one
variable given
as a one-dimensional integral.
\item Solve (trivially) $F_{1,\mathrm{ren}}=0$ (Eq.\ (\ref{F1eq0})) for $\beta$.
\item With $\{A=a/\beta, \mu=\nu/\beta,\kappa,T=1/\beta\}$ given, the potential density (\ref{24i}) and all observables
are given at one
point in the crystal phase in the $(T,\mu)$-plane.
\end{enumerate}
Note that the translationally invariant solution is obtained by setting $\kappa=1$ and skipping step 2.\par
\begin{figure}[h]
\begin{center}
\begin{psfrags}
\psfrag{mu}{$\mu$}
\psfrag{T}{$\!\!\!\!T$}
\epsfig{file=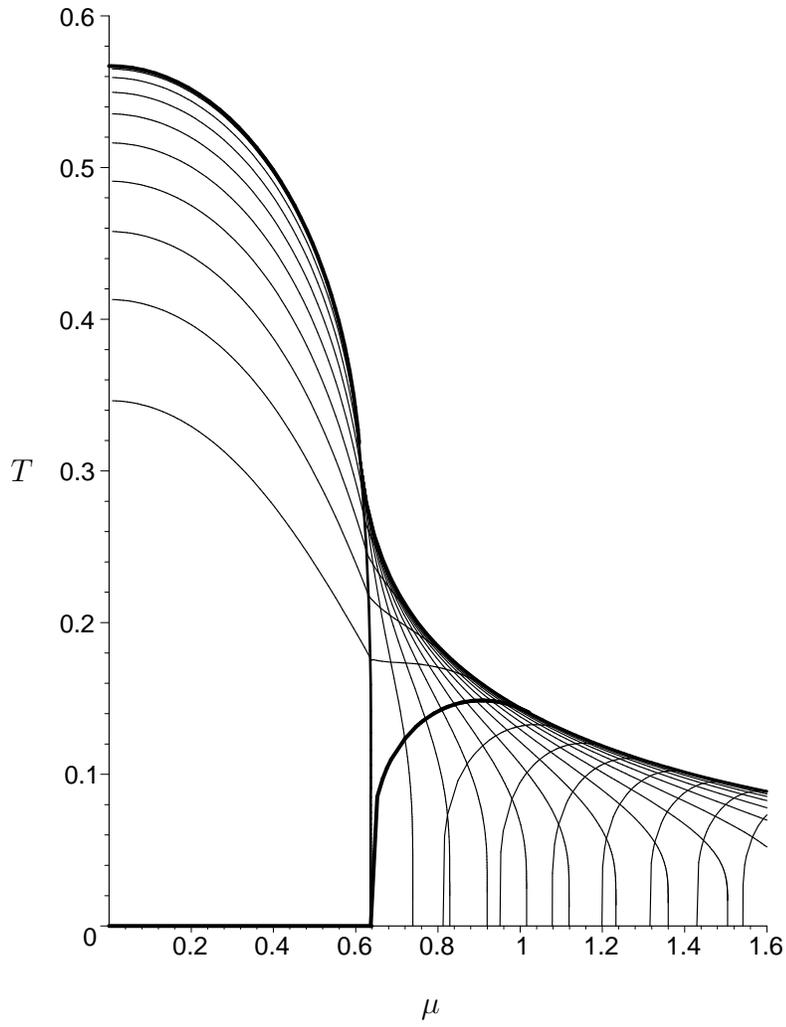,width=10cm}
\end{psfrags}
\caption{Lines of constant $A$ and $\kappa$ in the ($\mu,T$)-plane, obtained by
minimizing the grand potential. See Figs.\ \ref{FIG3} and \ref{FIG4} for more details.}
\label{FIG2}
\end{center}
\end{figure}
\begin{figure}[h]
\begin{center}
\begin{psfrags}
\psfrag{mu}{$\mu$}
\psfrag{T}{$\!\!\!\!T$}
\epsfig{file=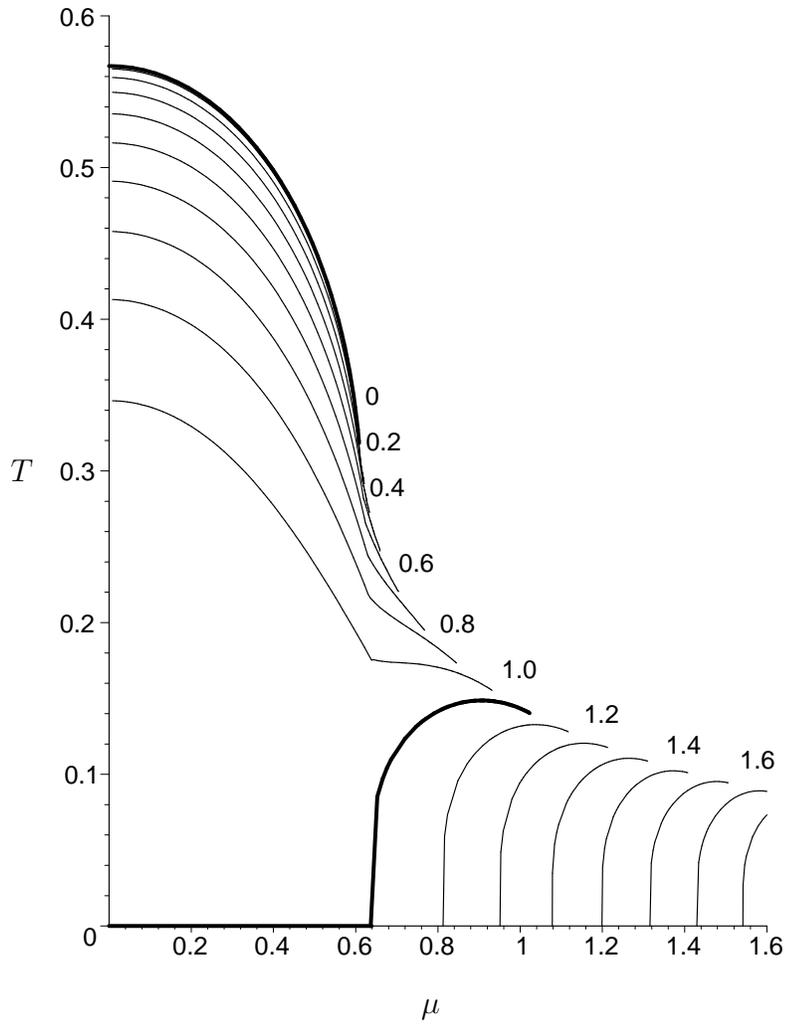,width=10cm}
\end{psfrags}
\caption{Contour lines $A=$ const. in the ($\mu,T$)-plane. Thick lines: $A=0$ and $A=1$. $A$ ranges from 0 to 1.7, the 
line spacing
is $\Delta A=0.1$. }
\label{FIG3}
\end{center}
\end{figure}
\begin{figure}[h]
\begin{center}
\begin{psfrags}
\psfrag{mu}{$\mu$}
\psfrag{T}{$\!\!\!\!T$}
\psfrag{k=0}{$\kappa\!=\!0$}
\psfrag{k=1}{$\!\!\!\kappa\!=\!1$}
\epsfig{file=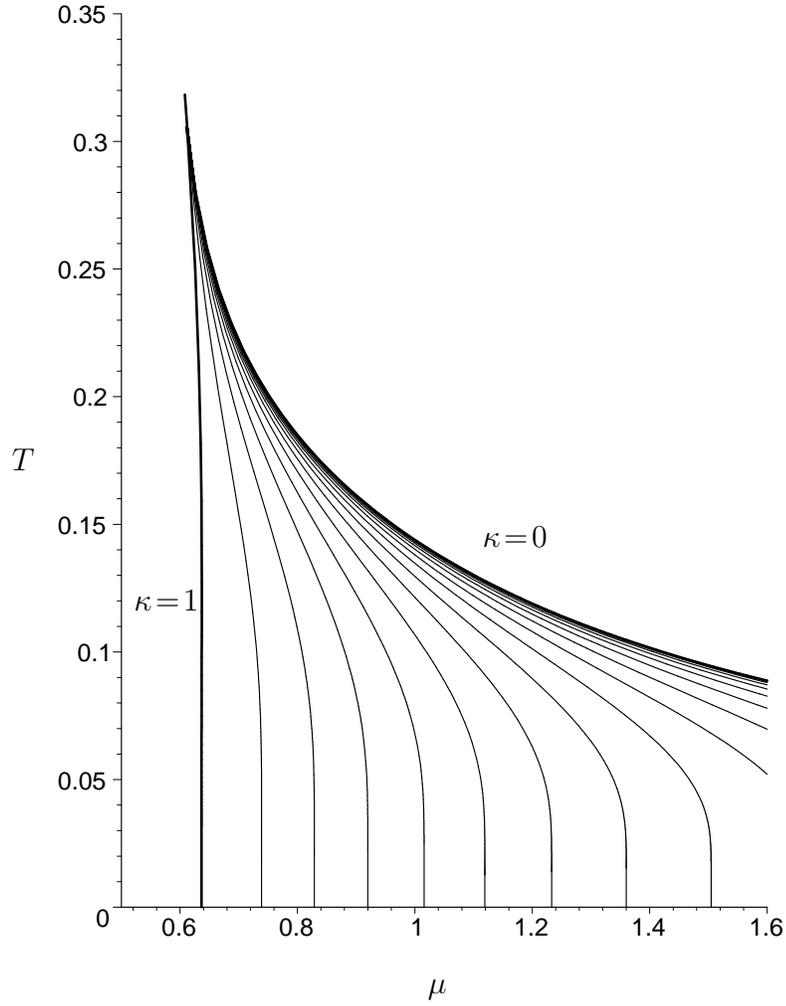,width=10cm}
\end{psfrags}
\caption{Contour lines $\kappa=$ const. in the crystal phase (closeup). The phase boundaries are $\kappa=0$ and 
$\kappa=1$, adjacent
lines differ by $\Delta \kappa =0.05$. All lines end at the tricritical point. At $T=0$ we have $A\kappa=1$ and
$\mu=2\mathbf{E}/\pi\kappa$.}
\label{FIG4}
\end{center}
\end{figure}

This enables us to compute the phase diagram and to explore all limits of interest (see Sect.\ \ref{section4}). We
illustrate the result
of the minimization procedure by plotting lines of constant $A$ and $\kappa$ in the ($\mu,T$)-plane, see Fig.\ \ref{FIG2}.
To clarify this
diagram, we also show separately the lines $A=$ const. and $\kappa=$ const. in Figs.\ \ref{FIG3} and \ref{FIG4}. The
thick lines in
Fig.\ \ref{FIG3} correspond to $A=0$ (phase boundary between massive and massless phases) and $A=1$. Like
a ``separatrix" the latter
curve divides the plot into regions where the lines $A=$ const. emanate from the $\mu$-axis (for $A>1$) and from
the $T$-axis (for $A<1$),
respectively. The other end of the curves lies on the phase boundary between crystal and massless phases. For $A<1$,
the kink visible
in the contour lines reflects the phase boundary between crystal and massive phases. In Fig.\ \ref{FIG4}, the thick curves
marked $\kappa=0$
and $\kappa=1$ correpond to phase boundaries between crystal phase and the two others. We shall return to this phase
diagram in
Sect.\ \ref{section4}. \par
For later use notice that the potential density at its minimum is equivalent to the simpler expression
\begin{eqnarray}
\pi\beta^2\Psi_1&\equiv& \pi\beta^2\Psi_{\mathrm{ren}}-\frac{a^2s^2}{2}F_1+a^2F_2
\label{24l} \\
&=& -\frac{\pi^2}{6}-\frac{\nu^2}{2}-\frac{a^2\kappa^2}{2\pi} \displaystyle \int\limits_0^{\pi/2}\mathrm{d}
\varphi \cos(2\varphi) \mathrm{Re}\left[\psi\left(\frac{1}{2}+\frac{i}{2\pi}(\nu+a\Delta_{\varphi})\right)+ (a \rightarrow -a)\right]
\nonumber
\end{eqnarray}

The standard thermodynamic observables can be computed in a straightforward manner, once the grand canonical potential
density $\Psi$ (see Eq.\ (\ref{24i})) is known. The pressure $P$, the (spatially averaged) baryon density $\rho$, the entropy
density $s$ and the energy density $u$ are given by
\begin{eqnarray}
P = - \Psi \, , \qquad & & \rho = -\frac{\partial}{\partial \mu} \Psi \ ,
\nonumber \\
s = \beta^2 \frac{\partial}{\partial \beta} \Psi \ , \qquad & & u = T s -P + \mu \rho \ .
\label{d9a}
\end{eqnarray}
The $x$-dependence of the baryon density in the crystal demands some extra attention. In analogy to
Eqs.\ (\ref{A40}), (\ref{A41}), we
find for $\rho(x)=\langle\psi^\dagger\psi\rangle_{\rm th}$
\begin{eqnarray}
\nonumber
\rho(x) & = & \frac{1}{\pi} \int\limits_0^{\Lambda/2} \mathrm{d}p \, \psi^\dagger \psi \left(\frac{1}
{\mathrm{e}^{\beta(\omega-\mu)}+1}
- \frac{1}{\mathrm{e}^{\beta(-\omega-\mu)}+1} \right) \\
& = & \frac{A}{\pi}\lim\limits_{\varepsilon\to 0 } \mathrm{Re} \int\limits_{-\frac{\Lambda\beta}{2a}+\mathrm{i}
\varepsilon}^{\infty+
\mathrm{i}\varepsilon}\mathrm{d}\omega\frac{\omega^2-1+\kappa^2/2+\tilde{S}^2/2}{\sqrt{(\omega^2-1+
\kappa^2)(\omega^2-1)}}\frac{1}{\mathrm{e}^{\beta(A\omega-\mu)}+1} \label{rho} \\
& = & \frac{\mu}{\pi}-\frac{A}{\pi^2} \int\limits_0^{\pi/2} \mathrm{d}\varphi \left(\Delta_{\varphi}+\frac{\tilde{S}^2-2
+\kappa^2}{2\Delta_{\varphi}}\right)\mathrm{Re} \left[ \psi\left(\frac{1}{2}+\frac{\mathrm{i\beta}}{2\pi}(\mu+
A\Delta_{\varphi})\right) -
(A\rightarrow -A)\right] . \nonumber
\end{eqnarray}
With Eqs.\ (\ref{Sspatialav}) and (\ref{24i}) we readily check that its spatial average satisfies $\langle \rho(x) \rangle
=-\partial\Psi/\partial \mu$.
We will evaluate $\rho(x)$ analytically at $T=0$ in the following section.


\section{Phase boundaries, zero temperature limit and multicritical point}
\label{section4}

{\bf Limit $\kappa \to 0$, perturbative phase boundary}\par

\begin{figure}[h]
\begin{center}
\begin{psfrags}
\psfrag{mu}{$\mu$}
\psfrag{T}{$T$}
\psfrag{k1}{\rotatebox{90}{$\!\!\!\!\!\!\!\kappa\!=\!1$}}
\psfrag{k0}{\rotatebox{-5}{$\kappa\!=\!0$}}
\psfrag{PL}{$P_{\rm L}$}
\psfrag{A0}{\rotatebox{-30}{$A\!=\!0$}}
\psfrag{crystal}{\fbox{crystal}}
\psfrag{meq0}{\fbox{$m=0$}}
\psfrag{mneq0}{\fbox{$m \neq 0$}}
\psfrag{EC}{$\frac{e^{\mathrm{C}}}{\pi}$}
\psfrag{2pi}{$2/\pi$}
\hskip-2.6cm \epsfig{file=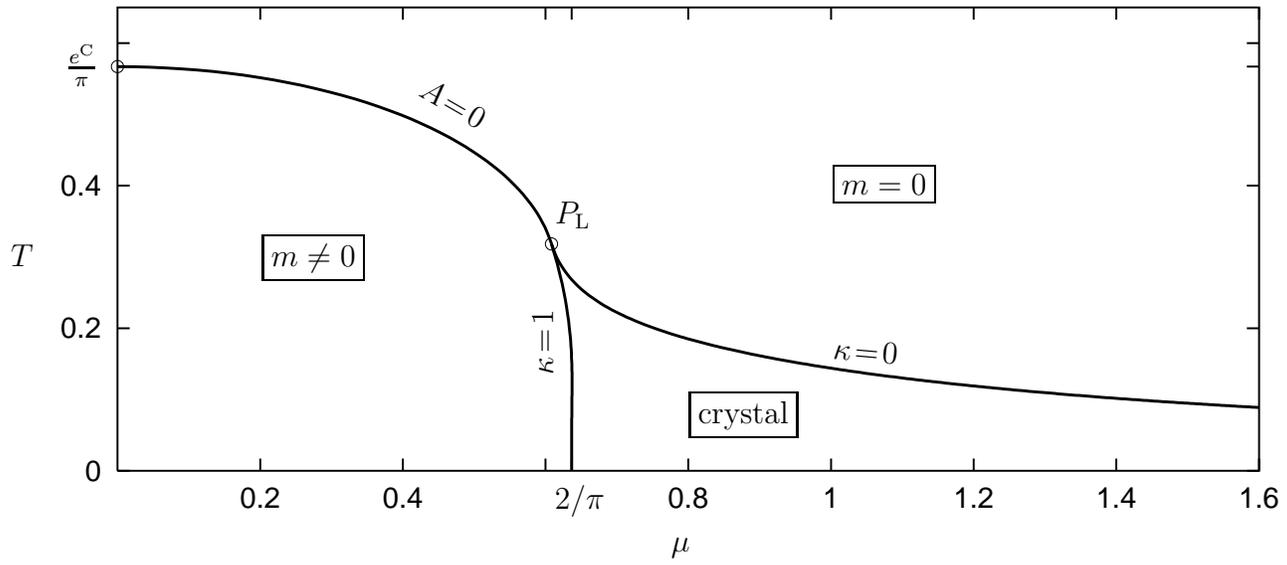,width=13.0cm, angle=-90}
\caption{Phase diagram of the GN model, computed with the help of Eqs.\ (\ref{phaseboundary1}), (\ref{phaseboundary1a}),
(\ref{C87a}),
(\ref{C87}). All phase boundaries correspond to second order transitions, $P_{\rm L}$ is the Lifshitz point 
determined by Eqs.\ (\ref{phaseboundary1a}), (\ref{w10}).}
\label{FIG5}
\end{psfrags}
\end{center}
\end{figure}

\begin{figure}[h]
\begin{center}
\begin{psfrags}
\psfrag{mu}{$\mu$}
\psfrag{T}{$T$}
\psfrag{k1}{\rotatebox{90}{$\!\!\!\!\!\!\!\kappa\!=\!1$}}
\psfrag{k0}{\rotatebox{-5}{$\kappa\!=\!0$}}
\psfrag{PL}{$P_{\rm L}$}
\psfrag{A0}{\rotatebox{-30}{$A\!=\!0$}}
\psfrag{crystal}{\fbox{crystal}}
\psfrag{meq0}{\fbox{$m=0$}}
\psfrag{mneq0}{\fbox{$m \neq 0$}}
\psfrag{EC}{$e^{\mathrm{C}}/\pi$}
\psfrag{2pi}{$2/\pi$}
\epsfig{file=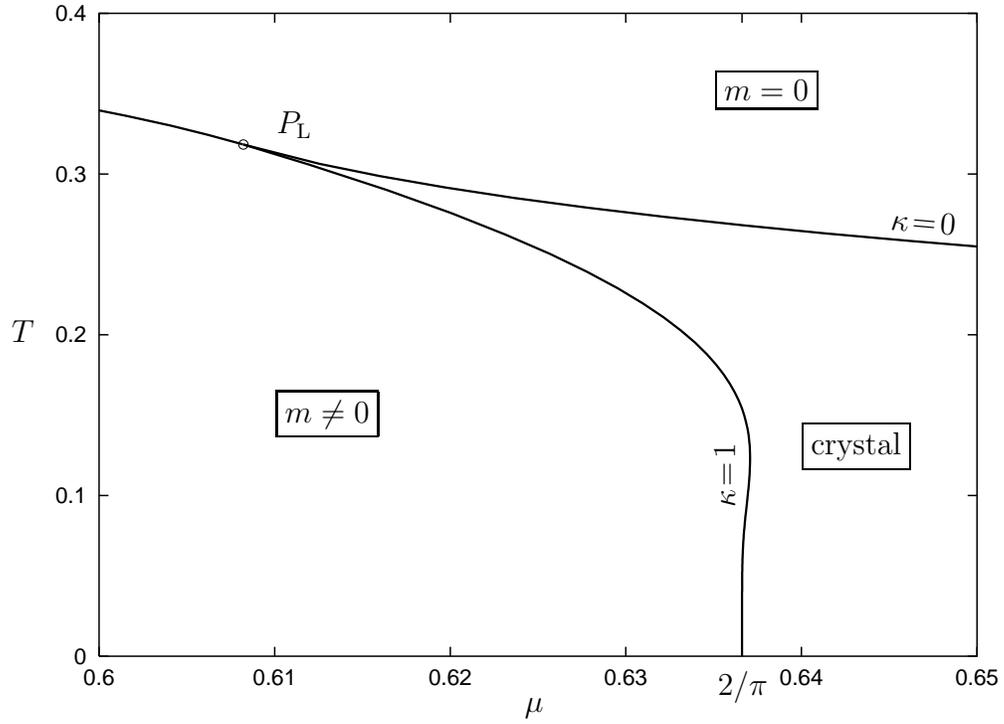,width=10cm,angle=-90}
\caption{More detailed view of the phase boundary between homogeneous and periodic ordered phases ($\kappa=1$).
Notice the
different scale on the $\mu$-axis as compared to Fig.\ \ref{FIG5}.}
\label{FIG6}
\end{psfrags}
\end{center}
\end{figure}

Here we are interested in the boundary between the chirally restored phase and the crystal. In Ref.\ \cite{16}
it has already been
determined via almost degenerate perturbation theory. As a cross-check we rederive it from the full thermodynamic
potential.
A straightforward expansion of $F_2$ [use Eq.\ (\ref{F2eq0})] yields
\begin{equation}
F_2 = - \frac{a\kappa^4}{32} \frac{\partial}{\partial a} \mathrm{Re} \left[\psi\left(\frac{1}{2}+\frac{\mathrm{i}}{2\pi}(\nu+a)\right) +
(a\rightarrow -a) \right] +\mathcal{O}(\kappa^6)\, . 
\end{equation}
With Eq.\ (\ref{F1eq0}) we can summarize the equations for the $\kappa\to 0$ phase boundary by
\begin{equation}
\label{phaseboundary1}
\ln\frac{1}{4\pi T}=\frac{1}{2}\min\limits_{a\geq 0}\mathrm{Re}\left[\psi\left(\frac{1}{2}+\frac{\mathrm{i}}{2\pi}(\nu+a)\right) + 
(a\rightarrow -a)\right]\, .
\end{equation}
The resulting curve is included in Figs.\ \ref{FIG5} and \ref{FIG6} (label ``$\kappa=0$"). For
small $\nu$ where $\mathrm{Re} \left[\psi\left(2,\frac{1}{2}+
\frac{\mathrm{i}\nu}{2\pi}\right)\right]<0$ ($\psi(2,x)=\mathrm{d}^2\psi(x)/\mathrm{d}x^2$ etc.), the unique minimum is
at $a=0$. In this
range of $\nu$ the phase boundary does not touch the crystal region in the $(\mu,T)$-diagram. It corresponds 
to the transition 
between the massless and the massive homogeneous solutions described by
\begin{equation}
\label{phaseboundary1a}
\ln\frac{1}{4\pi T} = \mathrm{Re} \, \psi\left(\frac{1}{2}+\frac{\mathrm{i\nu}}{2\pi}\right)\, .
\end{equation}
The critical temperature at $\mu=0$ (using $\psi(1/2)=- \mathrm{C}-\ln4$) coincides with the known value
\begin{equation}
T_c=\frac{\mathrm{e}^{\mathrm{C}}}{\pi} \, .
\end{equation}
The tricritical point is situated at $\nu_{t}$ with
\begin{equation}
\label{w10}
\mathrm{Re}\, \psi\left(2,\frac{1}{2}+\frac{\mathrm{i}\nu_{t}}{2\pi}\right) = 0\, , \qquad \nu_{t}=1.910668
\end{equation}
One finds $\mu_t=0.608221$, $T_t= 0.318329$, $\beta_t= 3.141401$ in agreement with the tricritical point of the old
phase diagram \cite{9}. \par
For $\nu>\nu_{t}$ the right hand side of Eq.\ (\ref{phaseboundary1}) develops a minimum for $a>0$, whereas $a=0$
becomes a maximum
(implying $T(a>0)>T(a=0)$). Here Eq.\ (\ref{phaseboundary1}) defines the boundary between the massless homogeneous
and the crystal
phase. In the neighbourhood of this boundary a leading order expansion allows us to analytically compare the values
of the minimized
potential density in the crystal phase with the massless homogeneous potential $\Psi(\kappa=0) = -\pi/6\beta^2-\mu^2/2\pi$.
To this end 
we fix a $\nu=\mu/T= \mathrm{const.}$ line in the $(\mu,T)$-diagram. This line has a unique intersection point 
$P_0=(\mu_0,T_0)$ with the 
phase boundary. The value $a_0$ of $a$ at $P_0$ is determined by the location of the minimum in Eq.\ (\ref{phaseboundary1}).
Now we determine the minimum value of $\Psi_{\mathrm{ren}}$ on the $\nu=\mathrm{const.}$ line in the neighbourhood 
of $P_0$ as a 
function of $T$ (and $\nu,a_0,T_0$) and compare it with the $\kappa=0$ result. We obtain
\begin{equation}
\Psi_{\mathrm{ren, min}}=\Psi_{\mathrm{ren},\kappa=0}+\frac{16\pi(T-T_0)^2}{\mathrm{Re}\left[\psi\left(2,\frac{1}{2}+
\frac{\mathrm{i}}
{2\pi}(\nu+a_0)\right) + (a_0\rightarrow -a_0)\right]} \, .
\end{equation}
The minimum condition in Eq.\ (\ref{phaseboundary1}) implies $0<\mathrm{d}^2(\mbox{r.h.s})/\mathrm{d}a^2$ at $a=a_0$ 
and thus 
$\Psi_{\mathrm{ren,min}}<\Psi_{\mathrm{ren,}\kappa=0}$ for $T<T_0$, $\nu>\nu_t$. The crystal solution is thermodynamically
favorable. 
This proves that a phase transition occurs. The $(T-T_0)$-dependence shows that it is of second order.

{\bf Limit $\kappa \to 1$, non-perturbative phase boundary}\par

This limit is relevant for the phase transition between crystal and massive phases as well as for the translationally 
unbroken, chirally 
broken ``massive'' phase alone. \par
The relation between $a$ and $\nu$ along the (non-perturbative) phase boundary is determined by $F_2=0$ or equivalently
\begin{equation}
\label{C87a}
\frac{1}{4a} \lim\limits_{\kappa \to 1} \ln\left(1-\kappa^2\right) F_2(\kappa) = \frac{\partial}{\partial a}\frac{1}
{a}\int\limits_0^{\pi/2}
\mathrm{d}\varphi \frac{1}{\cos\varphi} \mathrm{Im} \ln \frac{\Gamma\left(\frac{1}{2}+\frac{i}{2\pi}(\nu+a\cos \varphi)\right)}
{\Gamma\left(\frac{1}{2}+\frac{i}{2\pi}(\nu-a\cos \varphi)\right)} =0 \, .
\end{equation}
With these values for $a$ and $\nu$
\begin{equation}
\label{C87}
F_1(\kappa=1) =-\ln\frac{\beta}{4\pi}+\frac{1}{\pi}\int_0^\pi\mathrm{d}\varphi \, \mathrm{Re}\, \psi\left(\frac{1}{2}+
\frac{\mathrm{i}}{2\pi}(\nu
+a\cos\varphi)\right) = 0
\end{equation}
determines $\beta$ at the phase boundary. The resulting curve is shown in the phase diagram in Fig.\ \ref{FIG5}
(label ``$\kappa=1$"). An 
expanded plot
which reveals more details about the shape of the curve is displayed in Fig.\ \ref{FIG6}. Eq.\ (\ref{C87}) alone with 
values of $(a,\nu)$
that are not restricted by Eq.\ (\ref{C87a}) gives the connection between $a$, $\nu$ and $\beta$ in the massive phase 
away from the
crystal region.\par
As in the case $\kappa\to 0$ we can perform an analytical near-boundary comparison between the crystal and the
massive phase.
The result is
\begin{equation}
\Psi_{\mathrm{ren, min}} = \Psi_{\mathrm{ren,}\kappa=1} + \frac{2a_0^2 T_0}{\pi}\frac{T-T_0}{\ln(T-T_0)}
\end{equation}
where the relation between $a_0$ and $\nu$ is given by Eq.\ (\ref{C87a}). Since $T > T_0$, $\ln(T-T_0)<0$ in the crystal
phase near the
phase boundary, we have again established that the phase transition is genuine: In the region where mathematically
both phases 
exist, the crystal phase has lower potential. The $(T-T_0)$-dependence again shows that the
transition is second order.

{\bf Limit $T\to 0$}\par

Many equations simplify remarkably in the limit $T\to 0$, although the ansatz for $S(x)$ is basically unaltered. The
solution of the
$T=0$ case has been presented before \cite{19}. Here we want to show how to regain the $T=0$ results from the
general setting. \par
We start from Eqs.\ (\ref{F1eq0}), (\ref{F2eq0}) and (\ref{24l}) where we use asymptotic relations for large $z$: $\mathrm{Im}\ln
\Gamma(1/2+\mathrm{i}z) \sim z(\ln|z|-1)$, $\mathrm{Re}\, \psi(1/2 +\mathrm{i}z) \sim \ln|z|$ (here, $z=\beta(\mu\pm
A\Delta_\varphi)/2\pi$).
The integrals are best evaluated for $\mu>A$, where we do not have to account for the absolute value in the
logarithms. The
general result is recovered by taking the (unique) real value of the multivalued analytic continuation to $\mu<A$.
Standard techniques
(e.g. substitute $\Delta_\varphi=x$ and transform the integrals into contour integrals in the complex plane) yield
\begin{eqnarray}
F_{1,\mathrm{ren}}(T=0)& = &\mathrm{Re}\, \ln\left(\sqrt{\mu^2-A^2+A^2\kappa^2}+\sqrt{\mu^2-A^2}\,\right) \nonumber \\
F_2(T=0) & = & -\frac{\mathbf{E}}{\mathbf{K}}-\mathrm{Re}\,\frac{\mu }{A}\left[\left(1-\frac{\mathbf{E}}{\mathbf{K}}
\right)\mathbf{F}
\left(\frac{A}{\mu},\sqrt{1-\kappa^2}\right)-\mathbf{E}\left(\frac{A}{\mu},\sqrt{1-\kappa^2}\right)\right] \nonumber \\
\label{C102}
\Psi_1(T=0) & = & - \frac{\mu^2}{2\pi} +\frac{1}{4\pi}\mathrm{Re}\left[\sqrt{\mu^2-A^2+A^2\kappa^2}-\sqrt{\mu^2-A^2}\,\right]^2
\end{eqnarray}
where $\mathbf{F}/\mathbf{E}(x,\kappa)=\int_0^x (1-t^2)^{-1/2}(1-\kappa^2t^2)^{\mp 1/2}$ are the incomplete elliptic 
integrals of the first
and second kind.
For positive $A$ the only solution of $F_{1,\mathrm{ren}}=F_2=0$ falls into the range $A^2(1-\kappa^2)<\mu^2<A^2$ where 
Eqs.\ (\ref{C102}) simplify to (use $\mathbf{E}\mathbf{K}'+\mathbf{E}'\mathbf{K} - \mathbf{K}\mathbf{K}' = \pi/2$)
\begin{eqnarray}
F_{1,\mathrm{ren}}(T=0)& = & \ln (A\kappa) \nonumber \\
F_2(T=0) & = & -\frac{\mathbf{E}}{\mathbf{K}} + \frac{\mu}{A}\frac{\pi}{2\mathbf{K}} \nonumber \\
\Psi_1(T=0) & = & \frac{A^2 \kappa^2}{4\pi} - \frac{A^2}{2\pi}\, .
\end{eqnarray}
Solving $F_{1,\mathrm{ren}}=F_2=0$ gives 
\begin{equation}
A=\frac{1}{\kappa}
\end{equation}
and the following parameter representation of the grand canonical potential density,
\begin{equation}
\Psi_{\mathrm{ren}} = \frac{1}{4\pi}-\frac{1}{2\pi\kappa^2}\, ,\quad \mu =\frac{2\mathbf{E}}{\pi\kappa}\, . 
\end{equation}
We confirm that $A^2(1-\kappa^2)<\mu^2<A^2$ and find that $\Psi_{\mathrm{ren}}<\Psi_{\mathrm{ren}}(\kappa=1)$ and
$\Psi_{\mathrm{ren}}<\Psi_{\mathrm{ren}}(A=0)=-\mu^2/2\pi$. (For large $\mu$ we have $\Psi_{\mathrm{ren}}=
-\mu^2/2\pi-\mu^{-2}/64\pi$.)
This is consistent with the phase transition at $\mu=\mu(\kappa=1)=2/\pi$. \par
Moreover
\begin{eqnarray}
\rho (T=0) &=& - \frac{\partial}{\partial \mu}\psi \ =\ \frac{1}{2\kappa\mathbf{K}} \quad \mbox{ and} \nonumber \\
E_{\mathrm{ren}}(T=0) & = & \Psi_{\mathrm{ren}}+\mu\rho \ =\ \frac{1}{4\pi}+\frac{1}{\pi\kappa^2}\left(\frac{\mathbf{E}}
{\mathbf{K}}-
\frac{1}{2}\right) \, .
\end{eqnarray}
We recover the value of the Fermi momentum $p_{\mathrm{F}}=\pi \rho$ at $T=0$ from Ref. \cite{19}. \par
\begin{figure}[h]
\begin{center}
\begin{psfrags}
\psfrag{A=0}{a}
\psfrag{kappa=1 min}{b}
\psfrag{kappa=1 max}{c}
\psfrag{j=1}{d}
\psfrag{j=2}{e}
\psfrag{PSI}{$\Psi$}
\psfrag{MU}{$\mu$}
\psfrag{2pi}{$2/\pi$}
\epsfig{file=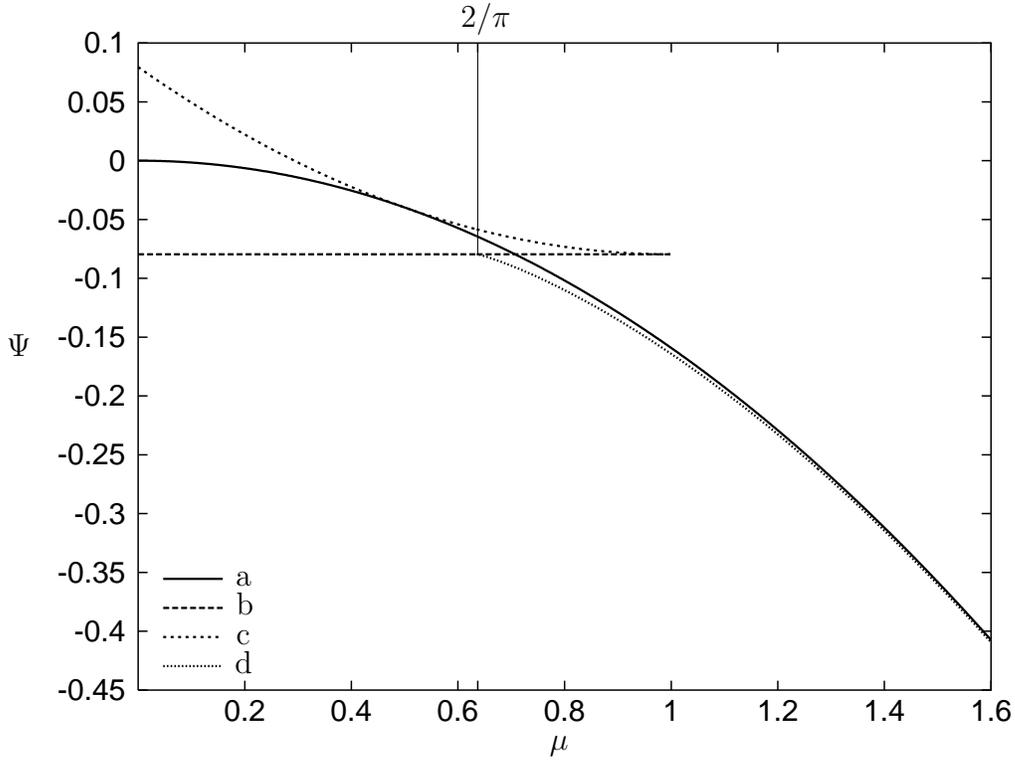,width=10cm,angle=-90}
\caption{The grand canonical potential density $\Psi$ for self-consistent solutions of the Dirac-Hartree-Fock 
equation at $T=0$:
(a) massless Fermi gas ($m=0$), (b) and (c) the minimum and maximum solution for the massive Fermi gas ($m=1$) 
and (d) the crystal
solution with the Lam\'e potential with single gap ($j=1$).
Notice that the crystal lies below the massless Fermi gas. At $\mu = 2/\pi$, a second order phase transition between 
the massive
Fermi gas and the crystal occurs.}
\label{FIG7}
\end{psfrags}
\end{center}
\end{figure}
\begin{figure}[h]
\begin{center}
\begin{psfrags}
\psfrag{A=0}{a}
\psfrag{kappa=1 min}{b}
\psfrag{kappa=1 max}{c}
\psfrag{j=1}{d}
\psfrag{j=2}{e}
\psfrag{PSI}{$\Psi$}
\psfrag{MU}{$\mu$}
\psfrag{2pi}{$2/\pi$}
\epsfig{file=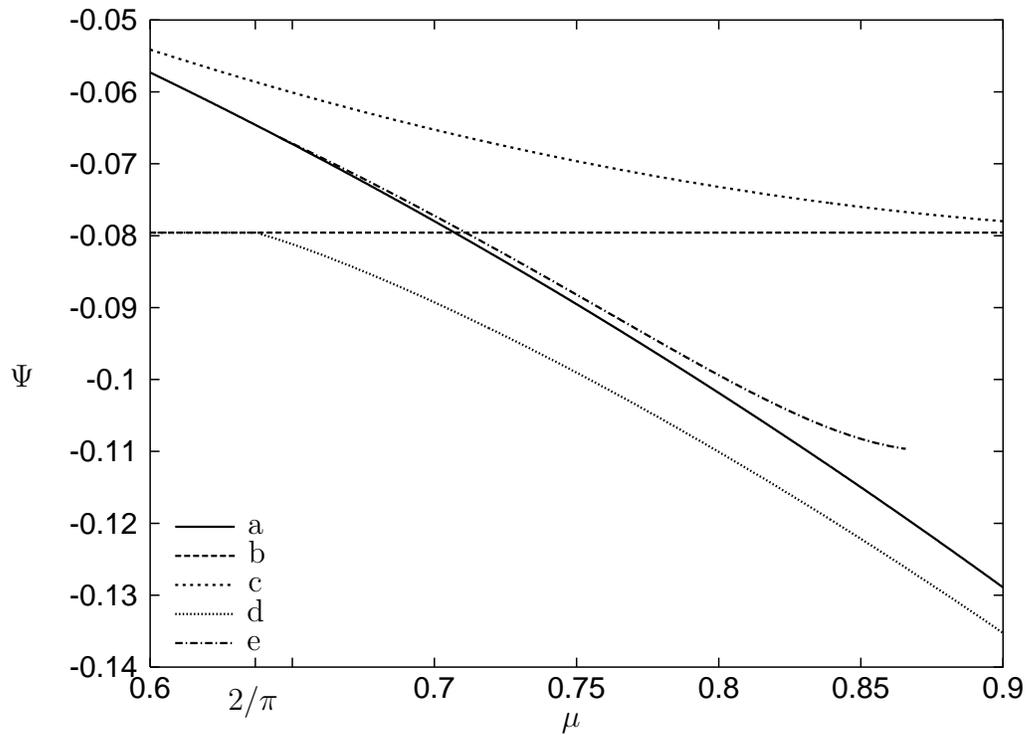,width=10cm,angle=-90}
\caption{A closeup of Fig.\ \ref{FIG7} showing $\Psi$ at $T=0$ for all known self-consistent Hartree-Fock solutions. 
Here the (unstable)
double gap Lam\'{e} solution ($j=2$) has also been included (e).}
\label{FIG8}
\end{psfrags}
\end{center}
\end{figure}

In Figs.\ \ref{FIG7} and \ref{FIG8}, $\Psi_{\mathrm{ren}}$ is plotted together with the self-consistent solution
of the homogeneous phases 
(two curves for 
the ``massive'' phase corresponding to the minimum and maximum of $\Psi_{\mathrm{ren}}$ and one curve for the 
``massless'' phase). \par
We end this section with a $T=0$ analysis of the $x$-dependent baryon density. From Eq.\ (\ref{rho})
we obtain
\begin{equation}
\rho(x,T=0) = \frac{1}{2\kappa\mathbf{K}} - \frac{\mathbf{K}'}{2\pi\kappa}\left(\tilde{S}^2(x/\kappa)-s^2\right) \, .
\end{equation}
This expression has the following high and low-density limits: At low density ($\kappa \to 1$) we recover the result
for a single baryon
\begin{equation}
\rho(x,T=0,\kappa \to 1) \approx \frac{1}{4\cosh^2x}+\frac{1}{4 \cosh^2(x/\kappa+\mathbf{K})} \, .
\end{equation}
At high density ($\kappa \to 0$) the total baryon density is constant
\begin{equation}
\rho(x,T=0,\kappa \to 0) \approx \frac{1}{2\kappa\mathbf{K}} \, .
\end{equation}

{\bf Limit $a \to 0$, perturbative phase boundary and tricritical point}\par

For all prominent functions in this paper the power series at $a=0$ can be given in closed form. With $c_n$ defined by
\begin{equation}
\frac{\omega^2}{\sqrt{(\omega^2-1+\kappa^2)(\omega^2-1)}} = \sum\limits_{n=0}^\infty c_n \omega^{-2n}
\end{equation}
we find $c_0=1$, $c_1=1-\kappa^2/2$, $c_2 = 1- \kappa^2 + 3\kappa^4/8$ and in general
\begin{equation}
\label{68}
c_n = \frac{2}{\pi} \int\limits_0^{\pi/2}\mathrm{d}\varphi \Delta_\varphi^{2n} = \sum\limits_{k=0}^n {2k\choose k} {n \choose k} 
\left(-\frac{\kappa^2}{4}\right)^k \, .
\end{equation}
We obtain
\begin{eqnarray}
F_{1,\mathrm{ren}} \! & = & \! -\ln \frac{\beta}{4\pi} + \sum\limits_{n=0}^\infty \left(-\frac{a^2}{4\pi^2}\right)^n\frac{c_n}{(2n)!}
\mathrm{Re}\,
\psi\left(2n,\frac{1}{2}+\frac{\mathrm{i}\nu}{2\pi}\right) \nonumber \\
F_2 \! & = & \! \sum\limits_{n=1}^\infty \left(-\frac{a^2}{4\pi^2}\right)^n \frac{\left((1-\kappa^2)c_{n-1}-\frac{\mathbf{E}}
{\mathbf{K}}
c_n\right)}{(2n-1)!(2n+1)}\mathrm{Re}\, \psi\left(2n,\frac{1}{2}+\frac{\mathrm{i}\nu}{2\pi}\right) \label{71}\\
\pi \beta^2\Psi_{\mathrm{ren}} \!& = & \! - \frac{\pi^2}{6} - \frac{\nu^2}{2} + a^2 \sum\limits_{n=0}^\infty \left(-\frac{a^2}{4\pi^2}\right)^n
\frac{c_{n+1}-\frac{\mathbf{E}}{\mathbf{K}}c_n}{(2n+1)!}\left[\mathrm{Re}\, \psi\left(2n,\frac{1}{2}+\frac{\mathrm{i}\nu}{2\pi}\right) 
- \delta_{n,0} \ln\frac{\beta}{4\pi} \right] . \nonumber
\end{eqnarray}
In the following we want to study the tricritical point by expanding the thermodynamical potential density to order
$a^6$. The result
from minimizing $\Psi_{\mathrm{ren}}$ with respect to $a$ and $\kappa$ is encoded in the fourth order expansions of
$F_{1,\mathrm{ren}}=F_2=0$. From $F_2=0$ we find
\begin{equation}
\label{C99}
\frac{a^2}{4\pi^2}=10\,G_1(\kappa^2)\frac{\mathrm{Re}\, \psi(2,\frac{1}{2}+\frac{\mathrm{i}\nu}{2\pi})}{\mathrm{Re}\, 
\psi(4,\frac{1}{2}+
\frac{\mathrm{i}\nu}{2\pi})}
\end{equation}
with
\begin{equation}
G_1(\kappa^2)=\frac{(1-\kappa^2)c_0-\frac{\mathbf{E}}{\mathbf{K}}c_1}{(1-\kappa^2)c_1-\frac{\mathbf{E}}{\mathbf{K}}c_2}
\end{equation}
and the limits
\begin{equation}
G_1(0) = \frac{3}{5}, \quad G_1(1)=\frac{4}{3} \, .
\end{equation}
If the right hand side of Eq.\ (\ref{C99}) is positive, there exists a non-trivial solution for $a$. The equation $F_1=0$
determines $\beta$ as function of $\kappa$ and $\nu$ to
\begin{equation}
\label{u4}
\ln\frac{\beta}{4\pi} = \mathrm{Re}\, \psi\left(\frac{1}{2} + \frac{\mathrm{i}\nu}{2\pi}\right) - \frac{5}{6}G_2(\kappa^2)\frac{\left
[\mathrm{Re}\,\psi (2,\frac{1}{2} + \frac{\mathrm{i}\nu}{2\pi})\right]^2}{\mathrm{Re}\,\psi (4,\frac{1}{2} + \frac{\mathrm{i}\nu}{2\pi})}
\end{equation}
with
\begin{equation}
\label{u6}
G_2(\kappa^2) = G_1(\kappa^2)\frac{(1-\kappa^2)(1-\kappa^2-\frac{3}{8}\kappa^4) - \frac{\mathbf{E}}{\mathbf{K}}c_1c_2}
{(1-\kappa^2)c_1
-\frac{\mathbf{E}}{\mathbf{K}}c_2}
\end{equation}
and the limits
\begin{equation}
G_2(0) = \frac{9}{5}, \quad G_2(1) = \frac{2}{3} \, .
\end{equation}
The potential density as function of $\kappa$ and $\nu$ is given by
\begin{equation}
\pi\beta^2\Psi_{\mathrm{ren}} = -\frac{\pi^2}{6} - \frac{\nu^2}{2}-\frac{50\pi^2\kappa^4}{3}G_1^2(\kappa^2)\frac{(1-\kappa^2)c_1
- \frac{\mathbf{E}}{\mathbf{K}} (1-\kappa^2+\frac{\kappa^4}{16})}{(1-\kappa^2)c_1 -\frac{\mathbf{E}}{\mathbf{K}}c_2}
\frac{\left[\mathrm{Re}\,\psi (2,\frac{1}{2} + \frac{\mathrm{i}\nu}{2\pi})\right]^3}{\left[\mathrm{Re}\, \psi (4,\frac{1}{2} +
\frac{\mathrm{i}\nu}{2\pi})\right]^2} \, .
\end{equation}
The two phase boundaries of the crystal phase correspond to $\kappa=0,1$. By using Eqs.\ (\ref{u4}) and (\ref{u6}), one can
plot $T=1/\beta$
against $\mu=\nu/\beta$ for $\nu \ge \nu_t $ and construct these phase boundaries near the Lifshitz point.
The
third, ``old'' phase boundary follows similarly from Eq.\ (\ref{phaseboundary1a}) for $\nu \le \nu_t$. Finally, in the
vicinity of the 
Lifshitz point, we can expand ($T,\mu$) around ($T_t,\mu_t$) and derive the (approximate) phase boundaries in 
closed analytical
form. Starting from Eq.\ (\ref{u4}), we find to second order
\begin{equation}
\ln \frac{\beta}{\beta_t} = - A_1 (\nu- \nu_t) - A_2 (\nu - \nu_t)^2
\label{u7a}
\end{equation}
with
\begin{eqnarray}
A_1 &=& \frac{1}{2 \pi} {\rm Im}\, \psi \left( 1, \frac{1}{2}+\frac{{\rm i}\nu_t}{2\pi}\right) \nonumber \\
A_2 & = & \frac{5}{24 \pi^2} G_2(\kappa^2) \frac{\left[ {\rm Im}\, \psi \left( 3, \frac{1}{2}+\frac{{\rm i}\nu_t}{2\pi}\right)
\right]^2}{{\rm Re}\,\psi\left( 4, \frac{1}{2}+\frac{{\rm i}\nu_t}{2\pi}\right)}
\label{u7b}
\end{eqnarray}
The equation for lines of constant $\kappa$ then reads
\begin{equation}
T-T_t = \frac{A_1 T_t}{\mu_t A_1+T_t }(\mu-\mu_t)- \frac{(A_1^2-2A_2)T_t^2}{2(\mu_tA_1+T_t)^3}
(\mu-\mu_t)^2\, .
\label{u7c}
\end{equation}
The linear term is independent of $\kappa$, so that all curves enter the Lifshitz point with the same 
slope. The quadratic term describes the $\kappa$-dependence of the curvature and changes sign
as a function of $\kappa$.
The limiting cases $\kappa=0,1$ yield the approximate phase boundaries in closed form. Furthermore, Eq.\ (\ref{C99}) shows 
that $a^2$ is linear in $\nu - \nu_t$. This translates into the behavior
\begin{equation}
A \sim (\mu- \mu_t)^{1/2}
\label{u7d}
\end{equation}
along the phase boundaries and the corresponding critical exponent $1/2$ as expected by mean field theory.\par
Finally, we derive a Ginzburg-Landau type effective action near the Lifshitz point from the Taylor expansion of the grand 
canonical potential in powers of $\kappa$. For this purpose, we compute spatial averages of powers of $S$ and its 
derivatives (keeping all even terms up to order $A^6$). Upon comparing the (analytical) results of such a calculation with 
Eqs.\ (\ref{68}), we find the simple relations
\begin{eqnarray}
c_1- \frac{\mathbf{E}}{\mathbf{K}} c_0 & = & \frac{s^2}{2}=\frac{1}{2A^2} \langle S^2 \rangle \nonumber \\
c_2 - \frac{\mathbf{E}}{\mathbf{K}} c_1 & = & \frac{3}{8A^4}\left(\langle S^4 \rangle +\langle (S')^2\rangle \right) \nonumber \\
c_3 - \frac{\mathbf{E}}{\mathbf{K}} c_2 & = & \frac{5}{16A^6} \left(\langle S^6\rangle +\frac{1}{2}\langle(S'')^2\rangle+
5\langle S^2(S')^2 \rangle \right)
\end{eqnarray}
This enables us to write down a Ginzburg-Landau effective action as follows,
\begin{eqnarray}
\Psi_{\rm eff} & = & - \frac{\pi}{6} T^2 - \frac{\mu^2}{2\pi} + \frac{1}{2\pi} S^2 
\left[ \ln (4\pi T)+ {\rm Re}\ \psi \left( \frac{1}{2} + \frac{{\rm i}\mu}{2\pi T}\right) \right]
\nonumber \\
& & -\frac{1}{2^6 \pi^3 T^2} \left( S^4 + (S')^2 \right) {\rm Re}\ 
\psi \left( 2, \frac{1}{2} + \frac{{\rm i}\mu}{2\pi T}\right) 
\label{w16} \\
& & + \frac{1}{2^{11}3\pi^5 T^4} \left( S^6 + \frac{1}{2} (S'')^2 +
5 S^2 (S')^2 \right) {\rm Re}\ \psi \left( 4, \frac{1}{2} + \frac{{\rm i}\mu}{2\pi T}\right) 
\nonumber
\end{eqnarray}
(This result can be applied to the translationally symmetric solution by simply dropping all the derivatives of $S$.) At the 
tricritical point, both the $S^2$ and $S^4+(S')^2$ terms vanish, see Eqs.\ (\ref{phaseboundary1a}) and (\ref{w10}). This 
is the reason why we had to include terms up to the order $S^6$. 

For completeness we report that we have found another self-consistent solution of the Dirac Hartree-Fock equation which 
corresponds to the $j=2$ (double gap) Lam\'{e} equation. The scalar density for this solution is
\begin{equation}
S_{j=2}(x) = 6 A \kappa^2 \frac{\mathrm{sn}(Ax,\kappa)\mathrm{cn}(Ax,\kappa)\mathrm{dn}(Ax,\kappa)}{1+\kappa^2+
\sqrt{1-\kappa^2+\kappa^4}
-3\kappa^2\mathrm{sn}^2(Ax,\kappa)}
\end{equation}
where again $A$ and $\kappa$ have to take specific values to ensure self-consistency for given $\mu$ and $T$. The solution
was found to be analytically accessible, being just slightly more involved than the $j=1$ case. However, the result showed
that (as physically expected from the gap-structure) the solution always lies higher in grand potential than the $j=1$ case.
Moreover, 
in general it does not correspond to a local minimum in the space of scalar potentials, having unstable 
directions. Therefore we refrain from writing down explicit equations. Instead we 
show in Fig.\ \ref{FIG8} a closeup of the $T=0$ plot (Fig.\ \ref{FIG7}) of the grand canonical potential 
density with the $j=2$ case included (in its range of existence $\frac{5}{12}<\mu^2<\frac{3}{4}$).


\section{Relation to condensed matter physics}
\label{section5}
In a previous paper, we have pointed out that GN type models can be reinterpreted as relativistic superconductors \cite{27}. 
This is due to a 2-dimensional remnant of the Pauli-G\"ursey symmetry of massless fermions \cite{27a,27b}. It allows
us to define
``particle'' and ``anti-particle'' independently for left-handed and right-handed quarks. Models with fermion-antifermion
pairing
(chiral condensate) and fermion-fermion pairing (Cooper pairs) can then be mapped onto each other by a ``duality'' 
transformation. 
The phase diagram which we have discussed above is equivalent to the phase diagram of a theory with Lagrangian
\begin{equation}
{\cal L} = \bar{\psi}^{(i)} {\rm i} \partial \!\!\!/ \psi^{(i)} + \frac{g^2}{2} \left( \psi^{(i)\dagger}_R \psi^{(i)\dagger}_L
+ \psi^{(i)}_L \psi^{(i)}_R \right)^2 \, ,
\label{se1}
\end{equation}
provided we replace the chemical potential $\mu=\mu_R+\mu_L$ by the
axial chemical potential $\mu_5 = \mu_R-\mu_L$ (see Ref.\ \cite{27}). The kink-antikink phase of the GN model
can then be identified with
the Larkin-Ovchinnikov-Fulde-Ferrel (LOFF) phase \cite{28,29} of the dual model. 
Such inhomogeneous superconductors have recently attracted considerable attention in the context of QCD (for a current
review article, see Ref.\ \cite{30}). 

Actually, there is yet another kind of connection between the GN model and non-relativis\-tic
condensed matter systems which is even more surprising. Although the GN model is mentioned
occasionally in the corresponding condensed matter literature (see e.g. \cite{30x,30y}), it seems that this close relationship
has never been exploited in a systematic way. The physics of the problems we have in mind is
totally different from the GN model,
yet the mathematical analogies are striking. Let us discuss a few pertinent references which we
came across when looking for parallels between the phase diagram of the GN model and that of
quasi-one-dimensional solid state systems. This may be regarded as a continuation of
the work of Jackiw and Schrieffer \cite{30a} on kinks in condensed matter and relativistic
field theories to finite kink densities.

The first example is a 1981 paper by Mertsching and Fischbeck on the Peierls-Fr\"ohlich model
\cite{31}. These authors address the quasi-one-dimensional Fr\"ohlich model with a nearly half-filled band,
an electron-phonon system. They are interested in the phase diagram, notably the transition between
commensurate-incommensurate charge density waves, using a mean field approximation.
There seems to be a mathematical one-to-one correspondence between this system and the 
GN model. The authors of Ref.\ \cite{31} have also found the analytic solution to the mean field equation, guided by the
Landau expansion around the triple point (which is called Leung point \cite{32} in this context). Crucial 
for the close correspondence with a relativistic field theory are evidently a continuum approximation 
(the lattice constant acts as inverse cutoff and is taken to 0) and a linearization of the electron dispersion
relation near the Fermi surface (simulating ``ultrarelativistic kinematics"). In appropriate units,
the results are identical to ours, as far as we can tell. 

In a subsequent paper, Machida and Nakanishi \cite{33} used the phase diagram of Ref.\ \cite{31} in a
different physics context: They studied the interplay of superconductivity and ferromagnetism
in ErRh$_4$B$_4$ (Erbium-Rhodium-Boride). Thanks to a number of approximations (in particular the
one-dimensional band model with linear dispersion) they managed to reduce this problem 
mathematically to the Peierls-Fr\"ohlich model. For real order parameter, their results are again
fully equivalent to ours for the GN model, except that now 
one has to use another dictionary: The Dirac equation corresponds to the Bogoliubov-deGennes equation,
particle/antiparticle degrees of freedom to spin (which exists in a quasi-one-dimensional world), chemical
potential to magnetic field, baryon density to spin polarization. Our three phases (massive,
crystal and massless) correspond to their BCS, ``sn" and normal phases, respectively. The order parameter at $T=0$
looks different from ours at first sight. However, the two expressions can be converted into each other
by Landen's transformation for Jacobi functions \cite{22} in the form
\begin{equation}
\kappa \frac{{\rm sn}(\xi,\kappa) {\rm cn}(\xi,\kappa)}{{\rm dn}(\xi,\kappa)}
= \frac{1-\kappa'}{\kappa} {\rm sn} \left( (1+\kappa')\xi, \frac{1-\kappa'}{1+\kappa'} \right) \, , \qquad \kappa'=\sqrt{1-\kappa^2} \, .
\label{w17}
\end{equation} 
Not only the phase boundaries, but all observables can be identified if one keeps in mind the above mentioned dictionary.
Nevertheless, there are subtle differences which show that one should not take over results blindly: The authors
of Ref.\ \cite{33} also discuss complex order parameters, in particular a helical phase. In the discrete chiral GN model,
the order parameter is always real.
In the continuous chiral NJL$_2$ model, it can be complex and we have indeed found a helical phase \cite{14},
but the results for the phase diagram are different. Here apparently the quantitative correspondence
between relativistic field theory and solid state physics ends due to some differences in dynamics.

As a third example, we would like to mention the more recent work of Buzdin and Kachkachi \cite{34}. They derive the 
Ginzburg-Landau 
theory for nonuniform LOFF superconductors near the tricritical point in the ($T,H$)-phase diagram in one, two
and three dimensions. If we take their result for one dimension and specialize it to a real order parameter, we 
find perfect agreement between our Eq.\ (\ref{w16}) and their Eq.\ (3) in appropriate units 
(a discrepancy in the sign of the $|\psi|^2$ term must be due to a misprint).
We have to identify their magnetic field ${\cal H}_0$ with our chemical potential $\mu$. 
One can find many more similar studies in the literature, see e.g. the recent review article on stripe phases
\cite{35}.

Our fourth example is related to solitons in trans-polyacetylene, starting with the seminal paper by Su, Schrieffer
and Heeger \cite{35a}. In this case there have been efforts to bridge the gap between particle and solid state physics
by Chodos and Minakata \cite{35b,35c}. Since these authors start from the old phase diagram of the Gross-Neveu model,
they notice that there are some discrepancies with the soliton lattice theory of Horovitz \cite{35d}. However, the
analytical results of Horovitz agree perfectly with our kink-antikink crystal so that everything falls into place.
Let us also mention in passing the results of Okuno and Onodera on the coexistence of a soliton and a polaron
in trans-polyacetylene \cite{35e} which are apparently closely related to the recent work of Feinberg \cite{35f}
on the analogous phenomenon in the Gross-Neveu model.

Finally, when browsing through the literature on superconductivity, we were amused to discover that the ``old'' phase 
diagram
of the GN model (assuming unbroken translational invariance) is ubiquitous in textbooks and articles on BCS theory 
(see e.g. Ref.\ \cite{36}, Fig.\ 6.2). It seems to go back to a 1963 paper \cite{37} which predates the original GN paper by more
than one decade and the first published GN phase diagram by more than two decades.

Summarizing, we find it gratifying that the GN model with its simple Lagrangian gives rise to such a physically rich 
phase diagram. 
Its structure seems to be mirrored by a variety of quasi-one-dimensional condensed matter systems. With hindsight, 
one might
ask the question why it took much longer to solve the same mathematical problem in relativistic quantum field theory 
than in 
condensed matter physics. We feel that this is due to the almost exclusive use of path integral methods in particle physics.
Apparently, these methods are not yet sufficiently developed to yield non-trivial saddle points with unexpected symmetry 
breakdown as in the present case. By contrast, canonical mean-field calculations offer a rich spectrum of approximation 
methods 
(variational, numerical etc.) which can be successively refined and used to nail down eventually the exact solution as 
in the case 
at hand. It is probably not accidental that we arrived at the phase diagram of the GN model using Hartree-Fock methods
akin to 
the standard tools of condensed matter physics. 

\bibliographystyle{unsrt}

\newpage

\newpage

\newpage

\newpage

\newpage

\newpage

\newpage

\newpage

\end{document}